\documentstyle[amstex,12pt,righttag]{article}
\newtheorem{theorem}{\indent Theorem}
\newtheorem{lemma}{Lemma}
\newtheorem{theorem*}{{}}
\newtheorem{claim}{Claim}

\def\th{{\theta}}
\def\thb{\overline{\theta}}
\def\D{{\cal D}}
\def\Db{\overline{\cal D}}
\def\l{\Lambda}
\def\lb{\overline{\Lambda}}

\def\cov{\nabla}

\def\mb{\overline{\mu}}
\def\nb{\overline{\nu}}
\def\rb{\overline{\rho}}
\def\sb{\overline{\sigma}}

\def\1b{\overline{1}}
\def\2b{\overline{2}}
\def\3b{\overline{3}}
\def\4b{\overline{4}}
\def\w{\wedge}
\begin{document}
\begin{flushright}
hep-th/9707012\\
Version 2.1
%Revised   17 July
\end{flushright}
\begin{center}
{\bf  {\Huge   D-Branes and  
Quotient    Singularities   \\
\vspace{0.5cm}
of  Calabi-Yau Fourfolds}} \\
\vspace{1cm}
 KENJI  MOHRI      \\
\vspace{0.5cm}
 {\it  Theory Group,                                   \\
Institute of Particle and Nuclear Studies,\\
High Energy Accelerator Research Organization (KEK) \\
 Oho 1-1 Tsukuba-city, Ibaraki  305-0801 Japan}\\
{e-mail}: {\tt mohri@@theory.kek.jp}\\
{ phone}: {\tt 81-298-64-5400}\\
{ fax}:   {\tt 81-298-64-5755}
\end{center}

\vspace{5cm}
\begin{abstract}
We investigate a (0,2) gauge theory realized on the world volume of
the type IIB D1-brane at the singular point of a Calabi-Yau fourfold.
It is argued that the gauge anomaly can be canceled via
coupling to the R-R chiral bosons in bulk IIB string.
We find that for a generic choice of the Fayet-Iliopoulos parameters
on the world volume, the Higgs moduli space is a smooth 
fourfold birational to the original Calabi-Yau fourfold, 
but  is not necessarily a Calabi-Yau manifold.
%
%
%
%%%%%%%%%%%%%%%%%%%%%%%%%%%%%%%%%%%%%%%%%%%%%%%%%%%%%%%%%%
    \vspace{1cm}                                        %
    \begin{flushleft}                                   %
    \end{flushleft}                                     %
%%%%%%%%%%%%%%%%%%%%%%%%%%%%%%%%%%%%%%%%%%%%%%%%%%%%%%%%%%
\end{abstract}
%
%---------------------------------------------------------
%
\newpage
\section{Introduction}
Recently the world volume gauge theory 
of the D-brane localized at the orbifold singularity of a Calabi-Yau 
3-fold \cite{DGM} has been formulated
generalizing the celebrated ALE case.\cite{DM,JM,BI}  
The Higgs moduli space in this theory can be   
identified with the effective spacetime that the D-brane probe feels
\cite{DKPS}
and is shown to be a smooth Calabi-Yau manifold which
is a resolution of the original orbifold.
The mathematical counterpart for construction of the moduli space
has been given in \cite{SI1,SI2}.
In this article we study the type IIB D1-brane gauge theory
localized at the orbifold singularity
of a Calabi-Yau 4-fold.
It is often the case  that the orbifold singularity of Calabi-Yau 4-folds
cannot be resolved without destroying Calabi-Yau property
(triviality of the 
canonical line bundle).\cite{SI3,Aspinwall,MS,Yukari,Reid2}

Thus it is quite interesting to verify whether
in these cases the D1-brane resolves the singularity of the orbifold
at the sacrifice the Calabi-Yau property 
or prefers the Calabi-Yau property rather than the smoothness
for  the effective spacetime.

We will show that the former possibility seems to be the case.
Moreover we will also show that 
the Higgs moduli space  is not necessarily a smooth Calabi-Yau 4-fold
even if the orbifold admits a smooth resolution of it.

\vspace{2cm}

The organization  of this article is as follows:

In section 2, we give general aspects of type IIB string
compactification  on  four dimensional Calabi-Yau orbifolds.
We also introduce three classes of singularity type there.
In section 3, the world volume (0,2) gauge theory of the D1-brane at 
the orbifold point is described using (0,2) superformalism.  
In section 4, we argue that the world volume gauge anomaly can be canceled
by the Chern-Simons coupling to the bulk chiral bosons.
Section 5 is devoted to the toric geometry of the Higgs moduli space.
In section 6, we give the toric description of the Higgs moduli space
for some examples.
In section 7, we give our conclusion and 
some remaining problems.
We present the (0,2) superformalism used in this article in section A 
and give an explicit form  of a gauge anomaly polynomial in section B.

\newpage

\section{Type IIB String on Orbifold}
\subsection{General Aspect}
%First we construct 
Let $\Gamma$ be a discrete subgroup of SU(4)
isomorphic to ${\Bbb Z}_{n}$.
The action $\pi$  of the generator $g$ on ${\Bbb C}^{4}$ is  
defined as follows,
\begin{equation}
\pi(g)=\begin{pmatrix}
\omega^{a_{1}}&      {}          &       {} &           {} \\
      {}      & \omega^{a_{2}}   &      {}  &           {} \\
 {}    &   {}   & \omega^{a_{3}}    & {}                 \\
{} & {} & {} & \omega^{a_{4}}
\end{pmatrix}
\end{equation}
where $\omega$ is a primitive $n$-th root of unity,
$\sum a_{\mu}\equiv 0 \mod{n}$ for the quotient space
$X_{\Gamma}:={\Bbb C}^{4}/\Gamma$ to have a nowhere vanishing
holomorphic 4-form
$\varOmega=\mbox{d}z^{1}\w \mbox{d}z^{2} \w \mbox{d}z^{3} \w \mbox{d}z^{4}$,
and  $(a_{\mu},n)=1$ for $X_{\Gamma}$
to have an isolated  singular point at the origin.\\
Note that the Lorentz quantum numbers of 
the 32 supercharges of type IIB string on $X_{\Gamma}$
are 
\begin{equation}
Q_{+}\otimes\left(1\oplus \Omega^{(0,2)}\oplus \Omega^{(0,4)}\right)
\oplus
Q_{-}\otimes\left( \Omega^{(0,1)}\oplus \Omega^{(0,3)}\right),
\end{equation}
where $Q_{+}$ and $Q_{-}$ are a right- and left-handed Dirac supercharges,
so that in general we have (0,4) supersymmetry in two dimensions
on  bulk closed string sector.

It is convenient   to
call  the quotient space $X_{\Gamma}$ above  
$\displaystyle{\frac{1}{n}(a_{1},a_{2},a_{3},a_{4})}$ model.

Let us present the bulk closed string spectrum on
$X_{\Gamma}$.
It is useful to assign the  number \cite{Aspinwall,Yukari,Reid2}  
$
\mbox{age}(k)\in \{ 1,2,3\},
$ 
to each twisted sector  $k$ 
with respect to a  primitive $n$-th root of unity $\omega$ by
\begin{equation}
\mbox{age}(k):=\sum_{\mu=1}^{4}\left[\negthickspace
\left[\frac{k a_{\mu}}{n}\right]\negthickspace\right],
\end{equation}
where
$ [\negthinspace[x]\negthinspace]:=x-[x]$ means  the fractional part of $x$.
It can be seen that the $k$-th twisted NS-NS sector 
of type II string on $X_{\Gamma}$ have a chiral primary field
corresponding to a generator of 
$H^{p,p}_{\mbox{\scriptsize phys}}(X_{\Gamma})$
for    $p=\mbox{age}(k)$.\cite{DFMS,Aspinwall}

Thus we can identify the physical Hodge number $h^{p,p}$
according to
\begin{equation}
h^{p,p}:=\#\left\{ k \big| 1\leq k \leq n-1,
\quad \mbox{age}(k)=p \right\}.
\end{equation}

In particular,
 each  twisted sector $k$  with $\mbox{age}(k)=1$
has  a moduli field $\phi_{k}$
the VEV of which induces a crepant blow-up of $X_{\Gamma}$;
a blow-up which preserves a Calabi-Yau property of
$X_{\Gamma}$. 
 The Euler number of $X_{\Gamma}$ is given by
$
e(X_{\Gamma})=1+h^{1,1}+h^{2,2}+h^{3,3}=n.      
$

On the other hand, each $k$-th twisted R-R sector with 
$\mbox{age}(k)=1,3$ ($\mbox{age}(k)=2$)
 gives rise to  a complex (anti-)self-dual 
scalar  $S_{k}$ ($A_{k}$) 
which satisfies 
$
\partial_{-}S_{k}=0, S_{k}^{*}=S_{n-k},
$
$
(\partial_{+}A_{k}=0,\ A_{k}^{*}=A_{n-k}).
$

The Kaluza-Klein origin of these fields \cite{Roy,DasM} is as follows:
D = 10  type IIB supergravity has a 2-form potential 
$C^{\left(2\right)}$ and  an anti-self-dual 4-form $C^{(4-)}$.
Contraction of $C^{(2)}$ by the $h^{1,1}$ harmonic 2-forms on
$X_{\Gamma}$ gives (non-chiral) $h^{1,1}$ scalars
in two dimensions.

The $h^{2,2}$  harmonic 4-forms
on $X_{\Gamma}$ are decomposed to the $(h^{2,2}-h^{1,1})$
self-dual 4-forms and the $h^{1,1}$ anti-self-dual ones
according to the Lefshetz decomposition.\cite{GH}
Thus contraction of $C^{(4-)}$ by the harmonic 4-forms
produces $(h^{2,2}-h^{1,1})$ anti-self-dual scalars, 
$\partial_{+}A=0$,
and $h^{1,1}$  self-dual scalars $\partial_{-}S=0$ in two dimensions.
                                                                               
Thus these Kaluza-Klein modes sum up to
 $(h^{2,2},2h^{1,1})$ chiral  scalars 
in one-to-one correspondence with those from the twisted
R-R sectors. 

These fields will play an important role in gauge anomaly cancellation 
on the D1-brane world volume.
\subsection{Resolvable Models}
Types of the singularity of $X_{\Gamma}$
are classified to the  following three classes,
which we call A,B and C for simplicity.

For physics literatures on
the toric geometry and the crepant resolutions of orbifold singularities,
see for example  \cite{AGM,Aspinwall}.

First we show the model that admits a crepant resolution,
%which means a resolution preserving a Calabi-Yau property
%of $X_{\Gamma}$,
just like the lower dimensional orbifold models
${\Bbb C}^{2}/\Gamma$, ${\Bbb C}^{3}/\Gamma$.
 
It seems that this class (we call A)  is relatively rare
and we know at present only two series plus one below:
\begin{claim}
$(1,1,1,3m-2)/(3m+1)$ model admits a crepant resolution.
\end{claim}
{\it Proof.}
The primitive generators of 1-cones of
the $(1,1,1,3m-2)/(3m+1)$ model is
\begin{eqnarray}
v_{1}&=&(3m+1,-1,-1,-(3m-2)),\nonumber \\
v_{2}&=& (0,1,0,0),          \nonumber \\
v_{3}&=& (0,0,1,0),                    \\
v_{4}&=& (0,0,0,1),           \nonumber 
\end{eqnarray}
and these four vectors constitute  the vertices 
of  a tetrahedron of volume $3m+1$,
which we call 
$\langle v_{1},v_{2},v_{3},v_{4}\rangle$.
There are  $m$ lattice points 
\begin{equation}
w_{i}=(i,0,0,-(i-1)), \quad 1\leq i \leq m,
\end{equation}
in the tetrahedron.
With these $m$ vectors we can triangulate (in a unique way)
the original tetrahedron into $3m+1$ tetrahedra of unit volume.\\
Thus the original $(1,1,1,3m-2)/(3m+1)$ model can be  resolved 
into a smooth Calabi-Yau 4-fold.\hfill $\Box$

The physical Hodge numbers of the $(1,1,1,3m-2)/(3m+1)$ model
is  $h^{1,1}=h^{2,2}=h^{3,3}=m$, 
which coincides with the Hodge numbers of
the resolved space.
\begin{claim}
$(1,1,2m-1,2m-1)/4m$ model admits a crepant resolution.
\end{claim}
{\it Proof.}
The primitive generators of 1-cones in this case are
\begin{eqnarray}
v_{1}&=&(4m,-1,-(2m-1),-(2m-1)),\nonumber \\
v_{2}&=& (0,1,0,0),          \nonumber \\
v_{3}&=& (0,0,1,0),                    \\
v_{4}&=& (0,0,0,1).           \nonumber 
\end{eqnarray}
First we see that the tetrahedron spanned by
$v_{1},..,v_{4}$ has a lattice point
$w_{0}:=(1,0,0,0)$ in it.
After the triangulation of the tetrahedron by $w_{0}$,
we have four tetrahedra
\begin{align*}
\sigma_{1}&=\langle w_{0},v_{2},v_{3},v_{4}\rangle,
\quad \mbox{vol}(\sigma_{1})=1,\\
\sigma_{2}&=\langle w_{0},v_{1},v_{3},v_{4}\rangle,
\quad \mbox{vol}(\sigma_{2})=1,\\
\sigma_{3}&=\langle w_{0},v_{1},v_{2},v_{4}\rangle,
\quad \mbox{vol}(\sigma_{3})=2m-1,\\
\sigma_{4}&=\langle w_{0},v_{1},v_{2},v_{3}\rangle,
\quad \mbox{vol}(\sigma_{4})=2m-1.
\end{align*}
Then there exist $(m-1)$ lattice points
$$
w_{i}=(2i+1,0,-i,-i), \quad i=1,..,m-1,
$$
inside the triangle 
$\sigma_{3}\cap \sigma_{4}=\langle w_{0},v_{1},v_{2} \rangle$.
The triangulation by these $(m-1)$ lattice points of the triangle
induces at the same time the triangulation of $\sigma_{3}$ and $\sigma_{4}$  
into tetrahedra of unit volume.
\hfill $\Box$

The physical Hodge numbers of $(1,1,2m-1,2m-1)/4m$ model are
$h^{1,1}=h^{3,3}=m$, $h^{2,2}=2m-1$.

The simplest example is $\displaystyle{\frac{1}{4}(1,1,1,1)}$ model,
 which can be blown-up to be a smooth Calabi-Yau 4-fold
(the canonical bundle of ${\Bbb P}^{3}$) ${\cal O}_{{\Bbb P}^{3}}(-4)$.

We also find an example $\displaystyle{\frac{1}{11}}(1,1,3,6)$
which is not contained in the above two series.

\subsection{Models with Terminal Singularity}
There are models which do not admit any crepant blow-up.
Such models are said to have a terminal singularity.
We call this class B.
In physical language,
compactification of type II string on these models 
don't yield  any moduli fields from  the twisted NS-NS sectors.
These models have already been  classified \cite{MS}:
\begin{theorem}(Morrison-Stevens)\\
$X_{\Gamma}$  has a  terminal singular point 
if and only if 
$(a_{1},a_{2},a_{3},a_{4})=(1,n-1,a,n-a)$
for a primitive $n$-th root of unity and  $(n,a)=1$.
\end{theorem}
The simplest example is $\displaystyle{\frac{1}{2}(1,1,1,1)}$ model.

We note that the above condition implies that  
the D1-brane world volume theory  has at least 
(0,4) supersymmetry because we have two extra covariant constant spinors
$\{\mbox{d}z^{\1b}\w \mbox{d}z^{\2b},\ 
\mbox{d}z^{\3b}\w\mbox{d}z^{\4b}\}.$ 

In these models,  each  twisted sector has age 2 so that 
the physical Hodge numbers are
$h^{1,1}=h^{3,3}=0, h^{2,2}=(n-1)$. 
\newpage
\subsection{Partially Resolvable Models}
The remaining  models which we call C class 
are characterized by the fact that they
admit some crepant blow-ups but not enough to
fully resolve singularities.
After the maximal crepant partial resolution,
there remain  terminal singularities in these models.
It seems that most of the $X_{\Gamma}$ belong to this class.

For example, take $\displaystyle{\frac{1}{5}(1,1,1,2)}$ model.

The primitive generators of 1-cones in this model are
\begin{eqnarray}
v_{1} &=& (5,-1,-1,-2)      \nonumber \\
v_{2} &=& (0,1,0,0)     \nonumber \\
v_{3} &=& (0,0,1,0)      \\
v_{4} &=& (0,0,0,1).     \nonumber
\end{eqnarray}
It is easily seen that the tetrahedron of volume 5 
defined by the above four vectors contains  only one lattice point
$w_{1}=(1,0,0,0)$. If we triangulate the tetrahedron incorporating
$w_{1}$, then we obtain three tetrahedra of unit volume and
one tetrahedron of volume two which cannot further subdivided into
two tetrahedra of unit volume as it doesn't have any lattice point in it.
Geometrically the partially resolved model has four open patches,
three of which is isomorphic to ${\Bbb C}^{4}$ and the remaining one 
has a terminal ${\Bbb Z}_{2}$ singularity. 

In general  the physical Hodge number $h^{1,1}$ coincides
with the number of the lattice points inside the tetrahedron
associated with $X_{\Gamma}$,
which have one-to-one correspondence with 
 crepant blow-ups of $X_{\Gamma}$. 
The lattice point $w_{k}$ 
inside the tetrahedron 
$\langle v_{1},v_{2},v_{3},v_{4}\rangle$
associated with the age 1 twisted sector $k$
is given by the useful formula:
\begin{equation}
w_{k}:=\sum_{\mu =1}^{4}\left[\negthickspace\left[ \frac{ ka^{\mu} }{n}
\right]\negthickspace\right]v_{\mu}.
\end{equation}

We also remark here that  any non-crepant blow-up
$Y\mapsto X_{\Gamma}$, in contrast  with  crepant ones,
changes the Euler number,
which can be calculated as the total volume of the
tetrahedra.

\newpage
\section{World Volume Theory}

\subsection{Orbifold Projection}
First we put $n$ D1-branes at the origin of ${\Bbb C}^{4}$
to obtain  a U$(n)$ super gauge theory on the world volume of the D1-branes.
The field content is as follows:\\
the gauge field $A_{\pm}$,
the four complex Higgs fields $Z^{\mu},\  \mu=1,..,4$,
where $(Z^{\mu})^{\dag}=Z^{\overline{\mu}}$,
the eight left-handed fermions,
\begin{equation}
 \Lambda\otimes \left(1\oplus \Omega^{(0,2)}\oplus \Omega^{(0,4)}\right)
=\Lambda\oplus \Lambda^{\mu\nu}\ \mbox{d}z^{\mb}\wedge \mbox{d}z^{\nb}
\oplus
\lb \
%\overline{\varOmega}\ 
\mbox{d}z^{\1b}\w \mbox{d}z^{\2b}\w \mbox{d}z^{\3b}\w \mbox{d}z^{\4b},
\end{equation}
where 
$(\l)^{\dag}=\lb$, \ 
$(\Lambda^{\mu\nu})^{\dag}=\frac{1}{2}\epsilon^{\mu\nu\rho\sigma}
\Lambda^{\rho\sigma}$,
and the eight right-handed fermions,
\begin{equation}
\Psi\otimes \left(\Omega^{(0,1)}\oplus \Omega^{(0,3)}\right)
=\oplus \Psi^{\mu}\ \mbox{d}z^{\mb}\oplus \epsilon^{\mu\nu\rho\sigma}
\Psi^{\mb}\ \mbox{d}z^{\nb}\w \mbox{d}z^{\rb}\w \mbox{d}z^{\sb},
\end{equation}
where $(\Psi^{\mu})^{\dag}=\Psi^{\mb}$. \\
Second we define the action of $\Gamma$ on the Chan-Paton factors 
by the regular representation:
\begin{equation}
\gamma (g)= \begin{pmatrix}
\omega      &        {}         &     {}    &       {}\\
   {}       &    \omega^{2}     &     {}    &       {} \\
   {}       &        {}         &  \ddots   &       {}  \\
   {}       &        {}         &     {}    & \omega^{n}
\end{pmatrix}.
 \end{equation}
Third to define a D1-brane on $X_{\Gamma}$, we implement
the orbifold projection by $\Gamma$ which
acts both on the Chan-Paton factors (via $\gamma$)
 and the Lorentz indices of ${\Bbb C}^{4}$ (via $\pi$);
\begin{eqnarray}
&A_{\pm}& = \gamma(g) A_{\pm}\gamma(g)^{-1},
\quad
\l      = \gamma(g) \l \gamma(g)^{-1},  \nonumber \\
&Z^{\mu}& = \omega^{a_{\mu}}\gamma(g) Z^{\mu} \gamma(g)^{-1},
\quad
\Psi^{\mu} = \omega^{a_{\mu}}\gamma(g) \Psi^{\mu} \gamma(g)^{-1} \\
&\l^{\mu\nu}&  = 
\omega^{a_{\mu}+a_{\nu}}\gamma(g) \l^{\mu\nu} \gamma(g)^{-1}. \nonumber
\end{eqnarray}
Then the  world volume fields of the D1-brane on the orbifold
$X_{\Gamma}$ are the matrix elements which survive 
the orbifold projection above,   
\begin{enumerate}
\item  
gauge field and left-handed Dirac fermion
\begin{equation}
A     =  (v_{i,i}),\quad
\l  = (\lambda_{i,i}), \label{vector-mult}
\end{equation}
\item
four pairs of complex bosons and right-handed Dirac fermions
\begin{equation}
Z^{\mu} = (z^{\mu}_{i,i+a_{\mu}}),\quad 
\Psi^{\mu} = (\psi^{\mu}_{i,i+a_{\mu}}),
\quad 1\leq \mu \leq 4, \label{bosonic-mult}
\end{equation}
\item
three left-handed Dirac fermions
\begin{equation}
\Lambda^{\mu\nu} = 
(\lambda^{\mu\nu}_{i,i+a_{\mu}+a_{\nu}}), 
\quad 1\leq \mu < \nu \leq 4. \label{fermionic-mult}
\end{equation}
\end{enumerate}
%
%%
%
%
%
%
%
%
%%%%%%%%%%%%%%%%%%%%%%%%%%%%%%%%%%%%%%%%%%%%%%%%%%%%%%%%%%%%%%%%%%%%%%%%%%
\newpage
\subsection{(0,2) Gauged Linear Sigma Model}
In this subsection we identify the world volume theory as 
a (0,2) gauged linear sigma model.
\cite{Phases,DK,Distler,KM}
For a recent development in this subject consult \cite{BSW}.  

It is natural to combine 
the fields in (\ref{vector-mult}) to the (0,2) vector multiplet
with gauge group $\mbox{U(1)}^{n}$,
and the fields in (\ref{bosonic-mult}) to
the bosonic chiral multiplet
$\boldsymbol\Phi^{\mu}$. \\
As for the three left-handed Dirac fermions in (\ref{fermionic-mult}),
there is an ambiguity in  assigning them (0,2) multiplet structure.

Here  we take a following route.
First set  $a_{4}=0$.
Then the world volume theory is the dimensional reduction 
of D = 4, N = 1 theory so that  we have  D = 2, (2,2) supersymmetry
on the D1-brane.

Under this (2,2) supersymmetry,
(0,2) vector multiplet and $\boldsymbol\Phi^{4}$,
which is diagonal when $a_{4}=0$, 
become a (2,2) vector multiplet,
while for $1\leq \mu \leq 3$,
$\boldsymbol\Phi^{\mu}$ and $ \Lambda^{4\mu}$
are combined into a (2,2) chiral multiplet $X^{\mu}$,
with the superpotential \cite{DGM}
\begin{equation}
W=\mbox{Tr}(X^{1}[ X^{2},X^{3}]). \label{(2,2)pot}
\end{equation}
Conversely the (0,2) reduction of the (2,2) formalism 
has been studied  in \cite{Phases}.
The result is simply that
$\Lambda^{\mu}:=\Lambda^{\mu 4}$ is in the (0,2) fermionic chiral multiplet
$\boldsymbol\Lambda^{\mu}$,
which satisfies
\begin{equation}
\Db \boldsymbol\Lambda^{\mu}
=\sqrt{2}[\boldsymbol\Phi^{\mu},\boldsymbol\Phi^{4}], \label{derivative}
\end{equation}
 and the remaining three fermions 
$\Lambda^{12}$,
$\Lambda^{23}$, and
$\Lambda^{31}$ are in the anti-chiral multiplets
${\boldsymbol\Lambda}^{\3b}$,
${\boldsymbol\Lambda}^{\1b}$,  and 
${\boldsymbol\Lambda}^{\2b}$ respectively.

Noting that (0,2) supersymmetry is unbroken even when
non-zero  $a_{4}$ is recovered,
we have  done the identification of the (0,2) model.
It will  become clear later that the physics doesn't depend on the choice of
the fourth dimension.

To sum up, the (0,2) gauge theory realized on the D1-brane world volume
localized at the origin of $X_{\Gamma}$ has the following field content:
\begin{enumerate}
\item 
$\mbox{U(1)}^{n}$ gauge multiplet the diagonal $\mbox{U(1)}$
of which is decoupled from the others 
\begin{equation}
\boldsymbol\Upsilon=\l-\th(F_{01}+iD)+2i\thb\th\partial_{+}\l,
\end{equation}
\item
four bosonic chiral multiplets,
%$\boldsymbol\Phi^{\mu}$, $1\leq \mu \leq 4,$
\begin{equation}
\boldsymbol\Phi^{\mu}=Z^{\mu}+\sqrt{2}\th \Psi^{\mu}
+2i\thb\th \cov_{+}Z^{\mu},
\end{equation}
\item
three fermionic chiral multiplets,
\begin{equation}
\boldsymbol\Lambda^{\mu}=\l^{\mu }-\sqrt{2}\th G^{\mu}
-2i\thb\th \cov_{+}\l^{\mu  }
-\sqrt{2}\thb[\boldsymbol\Phi^{\mu},\boldsymbol\Phi^{4}],
\end{equation}
\item
the (0,2) reduction of the superpotential (\ref{(2,2)pot}),
\begin{equation}
W=\sqrt{2}\mbox{Tr}\left(
\boldsymbol\Lambda^{1}[\boldsymbol\Phi^{2},\boldsymbol\Phi^{3}]+
\boldsymbol\Lambda^{2}[\boldsymbol\Phi^{3},\boldsymbol\Phi^{1}]+
\boldsymbol\Lambda^{3}[\boldsymbol\Phi^{1},\boldsymbol\Phi^{2}]
\right), \label{(0,2)pot}
\end{equation}
which satisfies $\Db W=0$.
\end{enumerate}

\newpage
Now it is straightforward to write down 
the total Lagrangian,\footnote{Here we are assuming 
flat metrics for the kinetic terms.}
\begin{equation}
L_{\mbox{\scriptsize tot}}=
L_{\mbox{\scriptsize gauge}}+
L_{\mbox{\scriptsize FI}}+
L_{\mbox{\scriptsize bos}}+
L_{\mbox{\scriptsize fer}}+
L_{\mbox{\scriptsize pot}}.
\end{equation}
The vector multiplet kinetic term is
\begin{equation}
L_{\mbox{\scriptsize gauge}}=
\frac{1}{2e^{2}}\int \mbox{d}\th \mbox{d}\thb
\mbox{Tr}\left(\overline{\boldsymbol\Upsilon}\boldsymbol\Upsilon\right)
= \frac{1}{2e^2}\mbox{Tr}
\left(
F_{01}^{2}+D^{2}+2i\lb\partial_{+}\l-2i\partial_{+} \lb \l
\right).
\end{equation}
The Fayet-Iliopoulos term is 
\begin{equation}
L_{\mbox{\scriptsize FI}}=
-\frac{1}{2}\int \mbox{d}\th \mbox{Tr}\left(\boldsymbol\Upsilon {\cal T}\right)
-\frac{1}{2}\int \mbox{d}\thb \mbox{Tr}
\left(\overline{\boldsymbol\Upsilon}\overline{\cal T}\right)=
\mbox{Tr}
\left( -RD+\frac{1}{2\pi}\Theta F_{01}
\right)\label{FI}
\end{equation}
where 
$
\displaystyle{{\cal T}=iR+\frac{1}{2\pi}\Theta}
$
 is a $n\times n$  constant diagonal matrix
which represents a Fayet-Iliopoulos parameter and
a theta parameter of $\mbox{U(1)}^{n}$ gauge theory.
We could assign any value we like to ${\cal T}$.  

The bosonic chiral multiplet kinetic term is
\begin{eqnarray}
L_{\mbox{\scriptsize bos}}&=&
i\int \mbox{d}\th \mbox{d}\thb
\mbox{Tr}
\left(
\boldsymbol\Phi^{\mb}\cov_{-}^{\mbox{\scriptsize s}}
\boldsymbol\Phi^{\mu}
  \right)\nonumber \\
&=&
2\mbox{Tr}\left(
\cov_{+}Z^{\mb}\cov_{-}Z^{\mu}
+\cov_{-}Z^{\mb}\cov_{+}Z^{\mu}+i\Psi^{\mb}\cov_{-}\Psi^{\mu}
 \right)\nonumber \\
&+&
\mbox{Tr}\left(
Z^{\mb}[D,Z^{\mu}]
-\sqrt{2}[Z^{\mb},\l]\Psi^{\mu}
+\sqrt{2}\Psi^{\mb}[\lb,Z^{\mu}]  \right)
\end{eqnarray}
The fermionic chiral multiplet kinetic term is
\begin{eqnarray}
L_{\mbox{\scriptsize fer}}&=&
\frac{1}{2}\int \mbox{d}\th \mbox{d}\thb
\mbox{Tr}\left(\boldsymbol\Lambda^{\mb}\boldsymbol\Lambda^{\mu}
\right)
=i\mbox{Tr}\left(
\l^{\mb}\cov_{+}\l^{\mu}-\cov_{+}\l^{\mb}\l^{\mu}\right)
\nonumber \\
&-&\sqrt{2}\mbox{Tr}\left(
([Z^{\4b},\Psi^{\mb}]-[Z^{\mb},\Psi^{\4b}]) \l^{\mu}
+\l^{\mb}([Z^{4},\Psi^{\mu}]-[Z^{\mu},\Psi^{4}])
\right)\nonumber \\
&+&
\mbox{Tr}\left(G^{\mb}G^{\mu}-2[Z^{4},Z^{\mu}][Z^{\mb},Z^{\4b}]\right).
\end{eqnarray}
Finally the superpotential term is
\begin{eqnarray}
L_{\mbox{\scriptsize pot}}
&=&-\frac{1}{\sqrt{2}}\int \mbox{d}\th W
-\frac{1}{\sqrt{2}}\int \mbox{d}\thb \overline{W}\nonumber \\
&=&
\sqrt{2}\mbox{Tr}
\left(
\l^{1}([Z^{2},\Psi^{3}]-[Z^{3},\Psi^{2}])+
([Z^{\2b},\Psi^{\3b}]-[Z^{\3b},\Psi^{\2b}])\l^{\1b}
\right)\nonumber \\
&+&
\sqrt{2}\mbox{Tr}
\left(
\l^{2}([Z^{3},\Psi^{1}]-[Z^{1},\Psi^{3}])+
([Z^{\3b},\Psi^{\1b}]-[Z^{\1b},\Psi^{\3b}])\l^{\2b}
\right)\nonumber \\
&+&
\sqrt{2}\mbox{Tr}
\left(
\l^{3}([Z^{1},\Psi^{2}]-[Z^{2},\Psi^{1}])+
([Z^{\1b},\Psi^{\2b}]-[Z^{\2b},\Psi^{\1b}])\l^{\3b}
\right)\nonumber \\
&+&
\sqrt{2}\mbox{Tr}
\left(
G^{1}[Z^{2},Z^{3}]- G^{\1b}[Z^{\2b},Z^{\3b}]
\right)\nonumber \\
&+&
\sqrt{2}\mbox{Tr}
\left(
G^{2}[Z^{3},Z^{1}]-G^{\2b}[Z^{\3b},Z^{\1b}]
\right) \nonumber \\
&+&
\sqrt{2}\mbox{Tr}
\left(
G^{3}[Z^{1},Z^{2}]-G^{\3b}[Z^{\1b},Z^{\2b}]
\right).   
\end{eqnarray}
We find the bosonic potential $U=U_{F}+U_{D}$, where
\begin{eqnarray}
U_{F}= 2\mbox{Tr}
\left(
[Z^{1},Z^{2}][Z^{\2b},Z^{\1b}] +
[Z^{2},Z^{3}][Z^{\3b},Z^{\2b}]+
[Z^{3},Z^{1}][Z^{\1b},Z^{\3b}]
\right.  \nonumber \\
+\left.
[Z^{1},Z^{4}][Z^{\4b},Z^{\1b}] +
[Z^{2},Z^{4}][Z^{\4b},Z^{\2b}]+
[Z^{3},Z^{4}][Z^{\4b},Z^{\3b}]
\right),
\end{eqnarray}
\begin{equation}
U_{D}= 
\frac{1}{2e^{2}}\sum_{i=1}^{n}D_{i,i}^{2}
=2e^{2}
\left[\mbox{Tr}
\left([Z^{\mu},Z^{\mb}]-R \right)^{2} %[Z^{\mb},Z^{\mu}]
\right].
\end{equation}
To be more explicit,
the $(\mu,\nu)$  term in F-term potential $U_{F}$  is
\begin{equation}
\mbox{Tr}
\left(
[Z^{\mu},Z^{\nu}][Z^{\nb},Z^{\mb}]
\right)
=\sum_{i=1}^{n}
 |z^{\mu}_{i,i+a_{\mu}}z^{\nu}_{i+a_{\mu},i+a_{\mu}+a_{\nu}}
 -z^{\nu}_{i,i+a_{\nu}}z^{\mu}_{i+a_{\nu},i+a_{\mu}+a_{\nu}}|^{2},
\end{equation}
and  the $i$-th D-term is written as 
\begin{equation}
D_{i,i}=-2e^{2}
\left(
 |z^{\mu}_{i,i+a_{\mu}}|^{2}
-|z^{\mu}_{i-a_{\mu},i}|^{2}
-r_{i}
\right).
\end{equation}
It is seen that for (0,2) supersymmetry to be unbroken,
the Fayet-Iliopoulos parameters $(r_{i})$ must be restricted 
to satisfy 
$\displaystyle{\sum_{i=1}^{n}r_{i}=0}$.
%

%%%%%%%%%%%%%%%%%%%%%%%%%%%%%%%%%%%%%%%%%%%%%%%%%%%%%%%%%%%%%%%%%%%%%%%
\newpage
\section{Gauge Anomaly on World Volume}
\subsection{Anomaly  Polynomial}
As the (0,2) gauge theory constructed above is chiral, we 
must check the consistency of the quantum theory.
It is convenient to
introduce the following $n\times n$ matrices
which encode the charge assignments of left- and right-handed
fermions:
\begin{alignat}{2}
(B_{R})_{i,j}:&= \sum_{\mu = 1}^{4}\delta_{j,i+a_{\mu}},
\quad
(C_{R})_{i,j}:&=4\delta_{i,j}-(B_{R})_{i,j}\\ 
(B_{L})_{i,j}:&=\sum_{\mu = 1}^{3}\delta_{j,i+a_{\mu}+a_{4}},
\quad
(C_{L})_{i,j}:&=3\delta_{i,j}-(B_{L})_{i,j}
\end{alignat}
Then the $\mbox{U(1)}^{n}$ gauge anomaly 4-form  of the theory 
can be written as 
\begin{equation}
{\cal A}
:=\left.
\bigoplus_{\mbox{\footnotesize right fermions }}
\mbox{Tr}(e^{F})
-\bigoplus_{\mbox{\footnotesize left fermions }}
\mbox{Tr}(e^{F})
\right|_{\mbox{\scriptsize 4-form}}
=\sum_{i,j=1}^{n}(C_{S})_{i,j}F_{i}F_{j}, \label{anomaly}
\end{equation}
where the real symmetric matrix $C_{S}$ is defined by
%\begin{equation}
$
C:= (C_{R})-(C_{L}), \ % 
C_{S}:= \frac{1}{2}(C+{}^{t}C).
%\end{equation}
$

We see that the gauge anomaly (\ref{anomaly})
does {\it not} vanish at all.

However the world volume theory 
of the D1-brane at the orbifold point should be a consistent theory,
thus there must exist some anomaly cancellation mechanism.

 We  show below that the interaction with the bulk closed string fields 
can solve the problem.
\subsection{Interaction with Bulk Closed String Fields}
The interaction term with NS-NS sector moduli fields
$\phi_{k}$ coming from the $k$-th twisted sectors
for  $\mbox{age}(k)=1$ 
as well as their anti-chiral  conjugates 
$\phi_{k}=\phi_{n-k}^{*}$ for $\mbox{age}(k)=3$ is
\begin{equation}
L_{\mbox{\scriptsize NS}}:=\int \mbox{d}^{2}x 
\sum
\begin{Sb}
k\\
{\mbox{\footnotesize age}(k)=1,3}
\end{Sb}
\phi_{k}
\mbox{Tr}\left(\gamma(g^{k})D \right),\label{NSNS}
\end{equation}
which might be absorbed into the original Fayet-Iliopoulos term on the 
world sheet (\ref{FI}).

On the other hand  we also have 
the Chern-Simons interaction with the chiral scalars 
in the R-R sector,\cite{DM,DGM}
\begin{align}
L_{\mbox{\scriptsize CS}}&:=
%\int %
%\sum_{k=1}^{n}C_{k}^{(0)}
%\frac{1}{\sqrt{n}}\mbox{Tr}\left(\gamma(g^{k})F \right)
%=\int \sum_{k=1}^{n}C_{k}^{(0)}\tilde{F}_{k},\nonumber \\
\int 
\sum
\begin{Sb}
k\\
\mbox{\footnotesize age}(k)=2
\end{Sb}
A_{k}\tilde{F}_{k}+
\sum
\begin{Sb}
k\\
\mbox{\footnotesize age}(k)=1
\end{Sb}
S_{k}\tilde{F}_{k}+
\sum
\begin{Sb}
k\\
\mbox{\footnotesize age}(k)=3
\end{Sb}
S_{k}\tilde{F}_{k} +C_{0}^{(0)}\tilde{F}_{0},\label{RR} \\
\tilde{F}_{k}&:=\frac{1}{\sqrt{n}}\sum_{i=1}^{n}\omega^{ik}F_{i},
\quad \tilde{F}_{n-k}=\tilde{F}_{k}^{*}.
\label{Fourier}
\end{align}
%
% 
%where $S_{k}$ ($A_{k}$) means the (anti-)self-dual scalar 
%in two dimensions coming from the $k$-th twisted sector
%which satisfies $S_{k}^{*}=S_{n-k}$ ($A_{k}^{*}=A_{n-k}$.)
%
Note that the diagonal $\mbox{U}(1)$ gauge field $\tilde{F}_{0}$
doesn't couple to the R-R chiral scalars
but to $C_{0}^{(0)}$, which is the R-R scalar of the $k=0$ sector.

\subsection{Cancellation of Anomaly}
The following is a crucial key to elucidate the structure of
${\cal A}$: 
\begin{lemma}
The $n$ eigenvectors $\{v_{k}\}$
and these eigenvalues $\{b_{k}\}$ of $C_{S}$ are given by  
\end{lemma}
\begin{align}
v_{k}:&={}^{t}\left(\omega^{k},\omega^{2k},..,\omega^{nk}\right),
\quad 1\leq k \leq n, \\
b_{k}=& (-1)^{\mbox{\footnotesize age}(k)}\cdot 8
\cdot \prod_{\mu=1}^{4}
\sin\left(\pi \left[\negthinspace\left[ka_{\mu}/n \right]
\negthinspace\right]\right).
%%%%%
\end{align}

%
%

%\end{lemma}

{\it Proof.}
The simple manipulation of the matrix shows that 
\begin{eqnarray*}
(2C_{S}v_{k})_{i}&=&
2(v_{k})_{i}
-\sum_{\mu=1}^{4}\left((v_{k})_{i+a_{\mu}}+(v_{k})_{i-a_{\mu}}\right)
+\sum_{\mu=1}^{3}\left((v_{k})_{i+a_{\mu}+a_{4}}
+(v_{k})_{i-a_{\mu}-a_{4}}\right)
 \\
&=& 2\left(1-\sum_{\mu=1}^{4}\cos\left(2\pi ka_{\mu}/n\right)
+\sum_{\mu=1}^{3}\cos\left(2\pi k(a_{\mu}+a_{4})/n\right)\right)(v_{k})_{i}.
%\qquad \Box
\end{eqnarray*}
Also by the direct calculation, we see that the eigenvalue above
is equal to 
\begin{equation*}
b_{k}=\frac{1}{2}\prod_{\mu=1}^{4}(\omega^{ka_{\mu}}-1)
=\frac{1}{2}\prod_{\mu=1}^{4}(e^{2\pi i[\negthinspace[ka_{\mu}/n
]\negthinspace]}-1).
\qquad\qquad \qquad  \Box
\end{equation*}

Note that $b_{n}=0$, $v_{n-k}=v_{k}^{*}$ and 
$v_{k}\cdot v_{l}^{*}=n\delta_{k,l}$.

%\begin{cor}
It follows that  as a quadratic form,
${\cal A}$  has the rank $(n-1)$ and the signature  $(h^{2,2},2h^{1,1})$.
More explicitly,
\begin{claim}
The anomaly polynomial ${\cal A}$ has the diagonalization: 
\end{claim}
\begin{equation}
{\cal A}=\sum_{k=1}^{n-1}b_{k}\tilde{F}_{k}\tilde{F}_{k}^{*}
=\sum
\begin{Sb}
k \\
{\mbox{\footnotesize age}(k)=2}
\end{Sb}
|b_{k}|\tilde{F}_{k}\tilde{F}_{k}^{*}
-\sum
\begin{Sb}
k \\
{\mbox{\footnotesize age}(k)=1}
\end{Sb}
|b_{k}|\tilde{F}_{k}\tilde{F}_{k}^{*}
-\sum
\begin{Sb}
k \\
{\mbox{\footnotesize age}(k)=3}
\end{Sb}
|b_{k}|\tilde{F}_{k}\tilde{F}_{k}^{*},
\end{equation}
{\it where $\tilde{F}_{k}$ is  the  complex linear combination of $F_{i}$
defined in (\ref{Fourier}).}

Now we see that the gauge anomaly of the world volume can be 
canceled by the counter term from the Chern-Simons interaction 
term (\ref{RR})
if an appropriate gauge transformation law \cite{Nep} is assigned 
to the $(h^{2,2},2h^{1,1})$ chiral bosons.

The anomaly cancellation mechanism described  
here may be regarded as a close analogue
of those in six dimensions. \cite{Sagnotti,BI}

\newpage
\section{Toric Description of Higgs Moduli Space}
%
%\subsection{Toric Description}
We denote the Higgs moduli space of the (0,2) gauge theory
as $M_{r}$, where the subscript $r$ reveals the Fayet-Iliopoulos parameter.
We construct $M_{r}$ in two steps following the procedure
described in \cite{DGM}.
A more abstract treatment  can be found in \cite{SI2}.

First let ${\cal N}$ to be the solution space of
the F-term constraint
\begin{equation}
 [Z^{\mu},Z^{\nu}]=0, \quad 1\leq \mu < \nu \leq 4,
\label{Fterm}
\end{equation}
which we will show to be a toric $(n+3)$-fold.\\
Second we take into account the D-term constraint
for $\mbox{U(1)}^{n-1}$ gauge group
\begin{equation}
\sum_{\mu=1}^{4}
\left( |z^{\mu}_{i,i+a_{\mu}}|^{2}-|z^{\mu}_{i-a_{\mu},i}|^{2}\right)
-r_{i}=0,
\quad  1\leq i \leq n-1. \label{Dterm}
\end{equation}
The quotient of ${\cal N}$ by the symplectic action of 
$\mbox{U(1)}^{n-1}$,  which is precisely the realm of
toric geometry, is the Higgs moduli space $M_{r}$,
\begin{equation*}
M_{r}:={\cal N}/\negmedspace/ \mbox{U(1)}^{n-1}.
\end{equation*}

\subsection{F-term Constraint}  
To solve the F-term constraint (\ref{Fterm}),
take the following generator of $\Gamma$, 
$g=(a,b,c,-1)$.\\
It is also useful to  make a following change of variables:
\begin{equation}
\left\{
\begin{gathered}
x_{i}:=z^{1}_{i,i+a},\\
y_{i}:=z^{2}_{i,i+b},\\
z_{i}:=z^{3}_{i,i+c},\\
w_{i}:=z^{4}_{i,i-1}
\end{gathered}\right.
\end{equation}
Using these variables, the constraints $[Z^{\mu},Z^{4}]=0$
can be written as
\begin{equation}
\left\{
\begin{gathered}
x_{i}w_{i+a}-w_{i}x_{i-1} = 0,  \\
y_{i}w_{i+b}-w_{i}y_{i-1} = 0,  \\
z_{i}w_{i+c}-w_{i}z_{i-1} = 0, 
\end{gathered}\right.
\end{equation}
from which we can solve for   $\{x_{i},y_{i},z_{i}\}$
in terms of $\{x_{0},y_{0},z_{0},w_{i}\}$ as follows:
\begin{equation}
\left\{
\begin{gathered}
x_{i}= \frac{w_{i}\cdots w_{1}\cdot w_{a}\cdots w_{1}}%
{w_{i+a}\cdot\cdots\cdots\cdots w_{1}}
x_{0},  \\
y_{i}= \frac{w_{i}\cdots w_{1}\cdot w_{b}\cdots w_{1}}%
{w_{i+b}\cdot\cdots\cdots\cdots w_{1}}
y_{0},            \\
z_{i}= \frac{w_{i}\cdots w_{1}\cdot w_{c}\cdots w_{1}}%
{w_{i+c}\cdot\cdots\cdots\cdots w_{1}}
z_{0}. \end{gathered}\right. \label{solution}
\end{equation}
It is easily seen that the solution (\ref{solution})
of $[Z^{\mu},Z^{4}]=0$ automatically satisfies 
the other constraints $[Z^{\mu},Z^{\nu}]=0$.
%%%%%%%%%%%%%%%%%%%%%%%%%%%%%%%%%%%%%%%%%%%%%%%%%%%%%%%%
%
\newpage
Thus ${\cal N}$ can be described by the $(n+3)$ variables
$\{x_{0},y_{0},z_{0},w_{0},..,w_{n-1}\}$. 
We also see that ${\cal N}$ is toric as it contains  
$({\Bbb C}^{*})^{n+3}$ as an open dense subset.  
The solution (\ref{solution}) as well as $w_{i}$s
 can be regarded as the
generators of  the affine coordinate ring $A_{\cal N}$ of ${\cal N}$,
which is translated to the cone $\sigma^{\vee}$ 
in the vector space ${\bold M}_{\Bbb R}
:={\bold M}\otimes_{\Bbb Z}{\Bbb R}
\cong {\Bbb R}^{n+3}$ as
$A_{\cal N}\cong \sigma^{\vee} \cap {\bold M}$,
where ${\bold M}\cong {\Bbb Z}^{n+3}$  is a lattice.\\
It is more appropriate to
realize ${\cal N}$ as a symplectic quotient
of ${\Bbb C}^{q}$ by $\mbox{U(1)}^{q-n-3}$ for some $q$.
For this purpose, let us define the dual cone 
$\sigma$ in ${\bold N}_{\Bbb R}$,
where ${\bold N}={\bold M}^{*}$, 
\begin{equation*}
\sigma:=\left\{ n\in {\bold N}_{\Bbb R}\left|\ \langle n , m\rangle\geq 0,
\quad {}^{\forall}m\in \sigma^{\vee}\right. \right\}.
\end{equation*}
The primitive generators of the 1-cones of $\sigma$:
$\left\{ v_{1},...,v_{q} \in {\bold N} \right\}$
 define the linear map
$
T:{\Bbb Z}^{q}\mapsto {\bold N}\cong {\Bbb Z}^{n+3},
$ 
the kernel of which 
gives  the $\mbox{U(1)}^{q-n-3}$
charge assignment to the homogeneous coordinates
$\left\{p_{1},p_{2},...,p_{q} \right\}$,
which is encoded in the $(q-n-3)\times q$ matrix $Q_{F}$.
The affine coordinates are recovered by 
\begin{equation*}
x_{m}:=\prod_{i=1}^{q}p_{i}^{\langle v_{i},m\rangle},
\ m\in \sigma^{\vee}\cap {\bold M}.
\end{equation*}

Thus we can express ${\cal N}$ as the following  symplectic quotient:
\begin{equation}
{\cal N}={\Bbb C}^{q}/\negmedspace/\mbox{U(1)}^{q-n-3}
\cong {\Bbb C}^{q}/({\Bbb C}^{*})^{q-n-3},
\end{equation}
where the ``Fayet-Iliopoulos parameters''
corresponding to $\mbox{U(1)}^{q-n-3}$ must set to be  zero
to obtain ${\cal N}$
because there are no excluded set in ${\Bbb C}^{q}$.

\subsection{D-term Constraint}
The original assignment  of the world volume gauge group 
$\mbox{U(1)}^{n}$ on the affine coordinate
$(z^{\mu}_{i,i+a_{\mu}})$ is simply  that
$(z^{\mu}_{i,i+a_{\mu}})$ has the $i$-th $\mbox{U(1)}$ charge +1
and the $(i+a_{\mu})$-th $\mbox{U(1)}$ charge ($-1$).
We encode the assignment of the world volume 
$\mbox{U(1)}^{n-1}$ charge
to the $q$ homogeneous coordinates of ${\cal N}$ in the 
$(n-1)\times q$ matrix $Q_{D}$.
Then we can unify the 
D-term constraint and F-term constraint by
 concatenation of  the two charge matrices 
$Q_{F}$ and $Q_{D}$ to the $(q-4)\times q$ matrix,
$
Q_{\mbox{\scriptsize tot}}:=
\begin{bmatrix}
\displaystyle{\frac{ Q_{F}}{Q_{D}}}
\end{bmatrix},
$
which defines the 
$\mbox{U(1)}^{q-4}$ charge assignment to the $q$ 
homogeneous coordinates $(p_{1},..,p_{q})$,
and we have the following symplectic quotient construction of
the Higgs moduli space:
\begin{equation}
M_{r}\cong 
{\cal N}/\negmedspace/\mbox{U(1)}^{n-1}
\cong {\Bbb C}^{q}/\negmedspace/\mbox{U(1)}^{q-4}. \label{Higgs}
\end{equation}
Thus we have realized $M_{r}$ as a toric 4-fold in (\ref{Higgs}).

The primitive 1-generators of  $M_{r}$
are denoted by $w_{i}\in {\Bbb Z}^{4}$ for later use.

Note that in the right hand side of (\ref{Higgs}),
only the last $(n-1)$ $\mbox{U(1)}$ charges,
which we call $(r_{1},...,r_{n-1})$ in (\ref{Dterm}), can  have non-zero
Fayet-Iliopoulos parameters.
%%%%%%%%%%%%%%%%%%%%%%%%%%%%%%%%%%%%%%%%%%%%%%%%%%%%%%%%%%%%%%%%%%%%%
\newpage
\section{Examples of Higgs Moduli Space}
The geometry of the Higgs moduli space $M_{r}$
is of great interest 
because it is the effective spacetime that the D1-brane probe feels.
There is a general statement about the geometry of 
the Higgs moduli space:\cite{SI1}
\begin{theorem}(Sardo Infirri)

If all Fayet-Iliopoulos parameters are zero, 
$M_{r=0}$ is isomorphic to
the original orbifold $ X_{\Gamma}$,
and if $r\ne 0$,  $M_{r}$
is always a (partial) resolution of $X_{\Gamma}$.
\end{theorem}
Here we investigate the cases where the $(n-1)$ Fayet-Iliopoulos parameters 
take generic values.

\subsection{A Model}
We  choose   $\displaystyle{ \frac{1}{4}}(1,1,1,1)$ model 
as the first example of  the A class.

The primitive generators $\{v_{1},..,v_{8}\in {\Bbb Z}^{7}\}$
of the 1-cones of $\sigma$ of the toric 7-fold ${\cal N}$
are as follows.
\begin{alignat*}{2}
v_{1}&=(1,0,0,0,0,0,0), &\quad
v_{2}&=(0,1,0,0,0,0,0),   \\
v_{3}&=(0,0,1,0,0,0,0), &\quad
v_{4}&=(0,0,0,0,1,0,0),   \\
v_{5}&=(0,0,0,0,0,1,0), & \quad 
v_{6}&=(0,0,0,0,0,0,1),     \\
v_{7}&=(1,1,1,1,0,0,0), & \quad
v_{8}&=(0,0,0,1,1,1,1).
\end{alignat*}

{}From these vectors we obtain the $1\times 8$  charge matrix
\begin{equation}
Q_{F}=
\begin{pmatrix}
1 & 1 & 1 & 1 & -1 & -1 & -1 & -1
\end{pmatrix},
\end{equation}
and the $3\times 8$  charge matrix for the world volume $\mbox{U(1)}^{3}$
charge assignment,
\begin{equation}
Q_{D}=
\begin{pmatrix}
0 & 0 & 0 & 0 & -1 &  1 &  0  & 0 \\
0 & 0 & 0 & 0 &  0 & -1 &  1  & 0 \\
1 & 1 & 1 & 1 & -1 & -1 & -2  & 0 
\end{pmatrix}.
\end{equation}

Concatenating the above two matrices, we have the following
toric realization of $M_{r}$:
\begin{equation}
\begin{matrix}
 p_{1} &  p_{2} &  p_{3} &  p_{4} &  p_{5} &  p_{6} &  p_{7} &  p_{8} 
&\mbox{\footnotesize FI}\\
  1    &    1   &    1   &   1    &  -1    &  -1    &   -1   &   -1   &  0\\
0    &    0   &    0   &   0    &  -1    &  1    &   0   &   0   & r_{1}\\
0    &    0   &    0   &   0    &  0    &  -1    &   1   &   0   & r_{2}\\
1    &    1   &    1   &   1    &  -1    &  -1    &  -2   &  0   & r_{3}.
\end{matrix}\label{toricdata1111/4}
\end{equation}
Furthermore by computing the kernel space of the $4\times 8$ charge matrix 
$
Q_{\mbox{\scriptsize tot}}=
\begin{bmatrix}
\displaystyle{\frac{ Q_{F}}{Q_{D}}}
\end{bmatrix},
$
we see the primitive generators $ w_{i}\in {\Bbb Z}^{4}$ of $M_{r}$
corresponding to $p_{i}$ above,
\begin{alignat*}{2}
w_{1}&=(-1,-1,-1,4), &\quad
w_{2}&=(1,0,0,0),   \\
w_{3}&=(0,1,0,0), & \quad
w_{4}&=(0,0,1,0), \\
w_{i}&=y_{0}:=(0,0,0,1),&\quad    5&\leq i \leq 8,
\end{alignat*}
which is just the toric data for the crepant resolution
of $X_{\Gamma}$
with $y_{0}$ the exceptional divisor.
Indeed for any generic values of $(r_{i})$,
or more precisely,  
for any $(r_{i})$ which satisfies 
$r_{i}\ne 0$, $r_{1}+r_{2}\ne 0$,
$r_{2}+r_{3}\ne 0$  and $r_{1}+r_{2}+r_{3}\ne 0$, 
$M_{r}$ is always the crepant resolution of $X_{\Gamma}$.

To see this, for example, take the combination of the charges as follows:
\begin{equation}
\begin{matrix}
 p_{1} &  p_{2} &  p_{3} &  p_{4} &  p_{5} &  p_{6} &  p_{7} &  p_{8} 
&\mbox{\footnotesize FI}\\
0    &    0   &    0   &   0    &  -1    &  1    &   0   &   0   &  r_{1}\\
0    &    0   &    0   &   0    &  -1    &  0    &   1   &   0   & 
r_{1}+r_{2}\\
0    &    0   &    0   &   0    &  -1   &   0    &   0   &   1   &
r_{1}+r_{2}+ r_{3}\\
1    &    1   &    1   &   1    & -4     &  0    &   0   &   0   &
3r_{1}+2r_{2}+r_{3}.
\end{matrix}\label{toricdata1111/4ex}
\end{equation}

Then we easily see that for $r_{1}>0$, $r_{1}+r_{2}>0$
and $r_{1}+r_{2}+r_{3}>0$, $\{p_{6},p_{7},p_{8}\}$
is decoupled and $M_{r}$ is the smooth Calabi-Yau 4-fold
which is the crepant blow-up of $X_{\Gamma}$.

Thus the effective geometry of the D1-brane on the singular point
of $X_{\Gamma}$  is the smooth resolution of it
 just like ${\Bbb C}^{2}/\Gamma$ models 
\cite{DM,JM} or ${\Bbb C}^{3}/\Gamma$ models.\cite{DGM}
  
%We have checked that this is also the case for
%$\displaystyle{\frac{1}{7}}(1,1,1,4)$.

%%%%%%%%%%%%%%%%%%%%%%%%%%%%%%%%%%%%%%%%%%%%%%%%%%%%%%%%%%%%%%%%%%%%

For the second example let us take 
$\displaystyle{\frac{1}{8}}(1,1,3,3)$ model.

The 26 primitive generators of 1-cones of the affine 
toric 11-fold ${\cal N}$
are given by 
\begin{alignat*}{2}
%%%%%%----      original 4 coordinates         ------------         
v_{1}&=(0,1,0,0,0,0,0,0,0,0,0),& \quad
v_{2}&=(1,0,0,0,0,0,0,0,0,0,0), \\
v_{3}&=(0,0,1,0,0,0,0,0,0,0,0),&  \quad
v_{4}&=(0,0,0,1,1,1,1,1,1,1,1), \\
%------------------- crepant divisor(I) -------------------------
v_{5}&=(0,0,0,0,1,0,1,0,0,1,0),& \quad
v_{6}&=(0,0,0,0,1,0,0,1,0,1,0), \\
v_{7}&=(0,0,0,0,1,0,0,1,0,0,1), & \quad
v_{8}&=(1,1,1,1,0,0,1,0,0,1,0), \\
v_{9}&=(0,0,0,0,0,1,0,0,1,0,1), & \quad
v_{10}&=(0,0,0,0,0,1,0,1,0,0,1), \\
v_{11}&=(0,0,1,1,0,0,1,0,1,0,0), & \quad
v_{12}&=(0,0,1,1,0,1,0,0,1,0,0), \\
%------------------ crepant divisor(II)   --------------------------
v_{13}&=(0,0,0,0,0,0,0,0,1,0,0),&  \quad
v_{14}&=(1,1,0,0,0,0,0,0,0,0,1), \\
v_{15}&=(1,1,1,1,0,0,0,0,0,0,0), & \quad
v_{16}&=(0,0,0,0,0,1,0,0,0,0,0), \\
v_{17}&=(1,1,0,0,0,0,0,0,0,1,0), & \quad
v_{18}&=(0,0,0,0,0,0,1,0,0,0,0), \\
v_{19}&=(0,0,0,0,1,0,0,0,0,0,0), & \quad
v_{20}&=(0,0,0,0,0,0,0,1,0,0,0), \\
%---------------------  non-crepant divisor
v_{21}&=(0,0,0,0,1,1,0,0,1,1,0), & \quad
v_{22}&=(1,1,1,1,0,1,0,1,0,1,0), \\
v_{23}&=(1,1,1,1,0,0,1,1,0,0,1), & \quad
v_{24}&=(0,0,1,1,1,0,0,1,1,0,0), \\
v_{25}&=(0,0,0,0,1,0,1,0,1,0,1), & \quad
v_{26}&=(1,1,0,0,0,1,1,0,0,1,1).
%----------------------------------------------
\end{alignat*}
Note that there is no vector 
$u\in {\bold M}\cong {\Bbb Z}^{11}$ such that
$u\cdot v_{i}=1, {}^{\forall}i$,
i.e.  ${\cal N}$ is not a Calabi-Yau variety.\cite{AGM}
\newpage
By  the standard procedure described in the previous section, 
we arrive at the toric data of $M_{r}$: 
\begin{align}
%----------- 
|p_{3}|^2+|p_{4}|^2-3|p_{5}|^2+|p_{13}|^2
&=-2r_{1}-r_{2}-3r_{3}-2r_{4}-2r_{6}-r_{7} \label{Q1} \\
|p_{1}|^2+|p_{2}|^2+|p_{5}|^2-3|p_{13}|^2
&=r_{1}+r_{2}+2r_{3}+2r_{4}-r_{5}   \label{Q2}     \\
-|p_{5}|^2-|p_{13}|^2+|p_{21}|^2
&=r_{2}      \label{Q3}  \\
%--------------- crepant divisor (I) --------------
|p_{6}|^{2}-|p_{5}|^{2}&=-r_{3} \label{Q56} \\
|p_{7}|^{2}-|p_{5}|^{2}&=-r_{3}-r_{6} \label{Q57} \\
|p_{8}|^{2}-|p_{5}|^{2}&=-r_{1}-r_{2}-r_{3}-r_{4}-r_{5}-r_{6}-r_{7}
\label{Q58} \\
|p_{9}|^{2}-|p_{5}|^{2}&=-r_{1}-r_{3}-r_{4}-r_{6}
\label{Q59} \\
|p_{10}|^{2}-|p_{5}|^{2}&=-r_{1}-r_{3}-r_{6}
\label{Q510} \\
|p_{11}|^{2}-|p_{5}|^{2}&=-r_{1}-r_{2}-r_{3}-r_{4}-r_{6}-r_{7}
\label{Q511} \\
|p_{12}|^{2}-|p_{5}|^{2}&=-r_{1}-r_{3}-r_{4}-r_{6}-r_{7}
\label{Q512} \\
%------------------- crepant divisor (II) ----------------------
|p_{14}|^{2}-|p_{13}|^{2}&=-r_{5}-r_{6}\label{Q1314}     \\
|p_{15}|^{2}-|p_{13}|^{2}&=-r_{5}-r_{6}-r_{7} \label{Q1315}      \\
|p_{16}|^{2}-|p_{13}|^{2}&=r_{2}+r_{3}+r_{4}    \label{Q1316}     \\
|p_{17}|^{2}-|p_{13}|^{2}&=-r_{5}        \label{Q1317}            \\
|p_{18}|^{2}-|p_{13}|^{2}&=r_{3}+r_{4}   \label{Q1318}             \\
|p_{19}|^{2}-|p_{13}|^{2}&=r_{1}+r_{2}+r_{3}+r_{4}  \label{Q1319}    \\
|p_{20}|^{2}-|p_{13}|^{2}&=r_{4}    \label{Q1320}          \\
%--------------------- non-crepant divisor ---------------------
|p_{22}|^{2}-|p_{21}|^{2}&=-r_{1}-r_{2}-r_{3}-r_{5}-r_{6}-r_{7}
\label{Q2122} \\
|p_{23}|^{2}-|p_{21}|^{2}&=-r_{1}-2r_{2}-r_{3}-r_{5}-2r_{6}-r_{7}
\label{Q2123}  \\
|p_{24}|^{2}-|p_{21}|^{2}&=-r_{2}-r_{3}-r_{6}-r_{7}
\label{Q2124}     \\
|p_{25}|^{2}-|p_{21}|^{2}&=-r_{2}-r_{6}
\label{Q2125}    \\
|p_{26}|^{2}-|p_{21}|^{2}&=-r_{1}-r_{2}-r_{5}-r_{6}. \label{Q2126}
\end{align}
We have also the primitive generators of the 1-cones of 
$M_{r}$:
\begin{align}
w_{1}&=(3,-4,0,-1),\quad
w_{2}=(0,0,0,1),\nonumber \\
w_{3}&=(-1,4,-1,0),\quad
w_{4}=(0,0,1,0), \\
w_{i}&=y_{0}:=(0,1,0,0),\quad 5\leq i \leq 12\nonumber \\
w_{i}&=y_{1}:=(1,-1,0,0),\quad 13\leq i \leq 20 \nonumber \\
w_{i}&=y_{2}:=(1,0,0,0), \quad 21\leq i \leq 26.\nonumber 
\end{align}
%
%%%%%%%%%%%%
We can make a following identification:
\begin{itemize}
\item
The first four vectors $w_{1},...,w_{4}$ are those of 
the original orbifold $X_{\Gamma}$.
\item
$y_{0}$ and  $y_{1}$ correspond to  the two crepant 
divisors of $X_{\Gamma}$.  
\item
$y_{2}(=y_{0}+y_{1})$ corresponds to a non-crepant divisor
incorporation of which destroys the Calabi-Yau property of $M_{r}$.
\end{itemize}
Note that the appearance of the non-crepant vector $y_{2}$ 
implies that  $M_{r}$ {\it may not be a Calabi-Yau manifold
despite of the existence of the crepant resolution.}

To show  this explicitly,
let $\tau_{1}$ and $\tau_{2}$ be 
the  adjacent 7-cones in the Fayet-Iliopoulos 
parameter space ${\Bbb R}^{7}$  defined by
\begin{align}
\tau_{1}:&=\mbox{cone}(
 \alpha_{1},\alpha_{2},\alpha_{3},\alpha_{4},
\alpha_{5},\alpha_{7},
\alpha_{8},\alpha_{9},\alpha_{10},\alpha_{11},\alpha_{12}),\\ 
\tau_{2}:&=\mbox{cone}( \alpha_{1},\alpha_{2},\alpha_{4},\alpha_{5},
\alpha_{6},\alpha_{7},\alpha_{9}), 
\end{align}
where the generators  of the cones are given by
\begin{alignat}{2}
\alpha_{1}&=(0,0,0,0,0,-1,1), &\quad
\alpha_{2}&=(0,0,0,0,0,0,-1),\nonumber \\
\alpha_{3}&=(0,1,0,0,0,-1,0), &\quad
\alpha_{4}&=(-1,0,0,1,0,0,0),\nonumber \\
\alpha_{5}&=(1,0,0,0,0,-1,0), &\quad
\alpha_{6}&=(0,-1,0,1,0,-1,0), \\
\alpha_{7}&=(0,0,-1,1,0,0,0), &\quad
\alpha_{8}&=(0,1,0,1,0,-1,-1),\nonumber \\
\alpha_{9}&=(0,0,0,0,-1,0,0), &\quad
\alpha_{10}&=(-1,1,0,0,0,-1,0),\nonumber \\
\alpha_{11}&=(1,1,0,0,-1,-1,-1), &\quad
\alpha_{12}&=(-1,1,-1,1,0,-1,-1).\nonumber 
\end{alignat}
{}From the toric data (\ref{Q1}-\ref{Q2126}), we see that  
for $(r_{i})\in \tau_{1}$, $M_{r}$ is the smooth Calabi-Yau  4-fold
which is the crepant resolution of $X_{\Gamma}$
with Euler number 8,
while for $(r_{i})\in \tau_{2}$, 
it is a  smooth non-Calabi-Yau 4-fold with Euler number 12.

However we have also observed an interesting fact:
if the Fayet -Iliopoulos parameters are restricted 
to those which come from
the VEV of the bulk moduli fields $\phi_{k}$ (\ref{NSNS}),
\begin{equation}
\left(r_{1},r_{2},r_{3},r_{4},r_{5},r_{6},r_{7}\right)=
\left(t_{1},t_{2},t_{3},t_{4},-t_{1},-t_{2},-t_{3}\right),
\end{equation}
then  $M_{r}$ is always the smooth Calabi-Yau 4-fold
for any generic choice of the parameters $(t_{1},t_{2},t_{3},t_{4})$.

%%%%%%%%%%%%%%%%%%%%%%%%%%%%%%%%%%%%%%%%%%%%%%%%%%%%%%%%%%%%%%%%%%%%%%%%%
%\newpage
\subsection{B Model}
We take $\displaystyle{\frac{1}{2}}(1,1,1,1)$ .
For a toric 5-fold ${\cal N}$,
we find the following six primitive generators:
\begin{alignat}{3}
v_{1}&= (1,0,0,0,0), &\quad 
v_{2}&= (0,0,1,0,0), &\quad  
v_{3}&= (0,0,0,0,1)\nonumber \\
v_{4}&= (0,0,0,1,1),&\quad 
v_{5}&= (0,1,0,0,0),&\quad
v_{6}&= (1,1,1,1,0).
\end{alignat}
{}From this data, we have 
\begin{equation}
Q_{F}=
\begin{pmatrix}
-1 & -1 & 1 & -1 & -1 & 1
\end{pmatrix},
\quad
Q_{B}=
\begin{pmatrix}
1 & 1 & -2 & 1 & 1 & 0
\end{pmatrix}.
\end{equation}
The charge assignment for the toric 4-fold $M_{r}$ is 
\begin{equation}
\begin{matrix}
 p_{1} &  p_{2}   &  p_{3} &  p_{4} &  p_{5} &  p_{6} & 
\mbox{\footnotesize FI}\\
  -1    &    -1   &    1   &   -1    &  -1    &  1  &      0    \\
  1    &      1   &   -2   &    1    &   1    &  0    &   r_{1}.  
\end{matrix}\label{toricdata1111/2}
\end{equation}
The primitive vectors $w_{i}$ corresponding to $p_{i}$ are
known from the kernel of $Q_{\mbox{\scriptsize tot}}$ as
\begin{alignat*}{3}
w_{1}&=(-1,-1,2,-1)&\quad
w_{2}&=(1,0,0,0) &\quad
w_{3}&=(0,0,1,0) \\
w_{4}&=(0,1,0,0)&\quad
w_{5}&=(0,0,0,1)&\quad
w_{6}&=(0,0,1,0),\nonumber
\end{alignat*}
where $w_{3}=w_{6}$ is identified with an exceptional divisor 
the incorporation of which resolves the singularity
of $X_{\Gamma}$ but at the same time destroys the Calabi-Yau property of it,
simply because $w_{3}=w_{6}$ is not inside the tetrahedron
(of volume 2) determined by the vertices of the other four vectors.

Indeed for $r_{1} > 0$, we can rewrite the toric data above 
(\ref{toricdata1111/2}) as
\begin{equation}
\begin{matrix}
 p_{1} &  p_{2}   &  p_{3} &  p_{4} &  p_{5} &  p_{6} &
 \mbox{\footnotesize FI}\\
  0    &    0   &    -1   &   0    &  0    &  1  &      r_{1}    \\
  1    &      1   &   -2   &    1    &   1    &  0    &   r_{1},  
\end{matrix}\label{toricdata1111/2plus}
\end{equation}
while for $r_{1} <0 $,
\begin{equation}
\begin{matrix}
 p_{1} &  p_{2}   &  p_{3} &  p_{4} &  p_{5} &  p_{6} & 
\mbox{\footnotesize FI}\\
  0    &    0   &    1   &   0    &  0    &  -1  &      -r_{1}    \\
  1    &      1   &   0   &    1    &   1    &  -2    &   -r_{1}.  
\end{matrix}\label{toricdata1111/2minus}
\end{equation}
{}From this  we see that {\it unless $r_{1}= 0$,
$M_{r}$ is a smooth  4-fold which is not Calabi-Yau.}

It is seen that $M_{r}$ has Euler number $4$.

%We also note that $\mbox{e}(M_{r})=4$
%as the resulting fan under  the triangulation by $w_{3}=w_{6}$
%consists with four tetrahedra of unit volume.

%This property also  exists in other members in the B class.

We here present the toric data of $M_{r}$ for 
$\displaystyle{\frac{1}{3}}(1,2,1,2)$ model as the second example.
\begin{equation}
\begin{matrix}
p_{3} & p_{4} & p_{5} & p_{10} & 
p_{1} & p_{6} & p_{7} & p_{2} & p_{8} & p_{9}&
\mbox{\footnotesize FI}\\
0  &  0  &  0  &  0  & -1 &  1  &  0  &  0  &  0  &  0 & r_{1} \\
0  &  0  &  0  &  0  &-1  &  0  &  1  &  0  &  0  &  0 & r_{1}+r_{2} \\
0  &  0  &  0  &  0  & 0  &  0  &  0  & -1  &  0  &  1 & -r_{2} \\
0  &  0  &  0  &  0  & 0  &  0  &  0  & -1  &  1  &  0 & r_{1} \\
1  &  1  &  2  &  2  & 0  &  0  &  0  & -3  &  0  &  0 & r_{1}-r_{2} \\
1  &  1  &  0  &  0  & -2  & 0  &  0  &  1  &  0  &  0 & r_{1}+r_{2}. 
\end{matrix}\label{toricdata1122/3}
\end{equation}
The primitive generators $w_{i}$ of the fan of $M_{r}$ is as follows:
\begin{align}
w_{3}&=(-1,2,-1,0),\quad
w_{4}=(0,0,1,0),\nonumber \\
w_{5}&=(2,-1,0,-1),\quad
w_{10}=(0,0,0,1), \nonumber \\
w_{i}&=y_{1}:=(0,1,0,0),\quad i=1,6,7 \\
w_{j}&=y_{2}:=(1,0,0,0), \quad j=2,8,9, \nonumber 
\end{align}
{}from which we can see that $y_{1}$ and $y_{2}$
correspond to  two exceptional divisors both of which are not crepant.

% and any generic choice of the Fayet-Iliopoulos
%parameters $(r_{1},r_{2})$ incorporates these two divisors in $M_{r}$.

It is easily seen 
in (\ref{toricdata1122/3}) that for $r_{1},r_{2}> 0$,
the coordinates $(p_{1},p_{9})$ survive and the resulting toric 4-fold
$M_{r}$ is not Calabi-Yau.
We can also check that as far as  $(r_{1},r_{2})$ takes generic value,
more precisely, $r_{1,2}\ne 0$ and $r_{1}+r_{2}\ne 0$,
the resulting toric 4-fold $M_{\Gamma}$ is the same 
(Ricci semi-negative) smooth
4-fold with Euler number $8$.
%$e(M_{r})=4$ for $\displaystyle{\frac{1}{2}}(1,1,1,1)$ and 
%$e(M_{r})=6$ for $\displaystyle{\frac{1}{3}}(1,2,1,2)$.

%%%%%%%%%%%%%%%%%%% C MODEL %%%%%%%%%%%%%%%%%%%%%%%%%%%%%%%%%%%%%
\subsection{C Model}
We take $\displaystyle{\frac{1}{5}}(1,1,1,2)$ model here.
The toric 8-fold ${\cal N}$ has 
14 primitive generators in its cone $\sigma$,
\begin{alignat*}{2}
v_{ 1} &=( 0, 0, 0, 0, 0, 1, 0, 0),& \quad 
v_{ 2} &=( 1, 1, 1, 1, 0, 0, 0, 0), \\
v_{ 3} &=( 0, 0, 0, 0, 1, 0, 0, 0), & \quad 
v_{ 4} &=( 1, 1, 1, 0, 0, 0, 1, 0), \\
v_{ 5} &=( 1, 1, 1, 0, 0, 0, 0, 1), & \quad 
v_{ 6} &=( 1, 0, 0, 0, 0, 0, 0, 0),  \\
v_{ 7} &=( 0, 1, 0, 0, 0, 0, 0, 0), & \quad 
v_{ 8} &=( 0, 0, 0, 1, 0, 1, 0, 0),           \\
v_{ 9} &=( 1, 1, 1, 1, 0, 0, 1, 0), & \quad 
v_{10} &=( 0, 0, 1, 0, 0, 0, 0, 0),  \\
v_{11} &=( 0, 0, 0, 0, 1, 0, 1, 0), & \quad 
v_{12} &=( 0, 0, 0, 0, 1, 0, 0, 1), \\
v_{13} &=( 0, 0, 0, 0, 0, 1, 0, 1), & \quad 
v_{14} &=( 0, 0, 0, 1, 1, 1, 1, 1),  
\end{alignat*}
{}from which we find the charge matrices $Q_{F}$, $Q_{D}$ as follows.
\begin{equation}
Q_{F}=
\begin{pmatrix}
-1 & -1 & 0 &  0 & 0 & 1 & 1 & 1 & 0 & 1 & 0 & 0 & 0 & 0 \\
 1 &  1 & -1&  -1& 0 & 0 & 0 & -1&0  & 0 & 1 & 0 & 0 & 0 \\
 1 &  0 & 0 & -1 & 0 & 0 & 0 & -1& 1 & 0 & 0 & 0 & 0 & 0 \\
 1 &  1 & -1&  0 &-1 & 0 & 0 & -1& 0 & 0 & 0 & 1 & 0 & 0 \\
 0 &  1 & 0 &  0 &-1 & 0 & 0 & -1& 0 & 0 & 0 & 0 & 1 & 0 \\
 2 &  2 & -1& -1 &-1 & 0 & 0 &-3 & 0 & 0 & 0 & 0 & 0 & 1 
\end{pmatrix}.
\end{equation}
\begin{equation}
Q_{D}=
\begin{pmatrix}
 0 & 0  & 0 &  0 & 0 & 1 & 1 & -1& -1& 1 & -1& -2& 0 & 2 \\
 0 & 0  & 0 &  0 & 0 & 0 & 0 & -1&  1& 0 & 0 & 0 & 0 & 0 \\
 0 &  0 & 0 &  0 & 0 & 0 & 0 & 0 & 0 & 0 & -1& 1 &0  & 0 \\
 0 & 0  & 0 &  0 & 0 & -1& -1& 2 & 1 &-1 & 1 & 1 & 0 &-2 
\end{pmatrix}.
\end{equation}
The 14 primitive generators $w_{i}\in {\Bbb Z}^{4}$
of the 1-cones of $M_{r}$ which correspond to the coordinates
$ p_{i}$ are computed from the kernel space of 
$Q_{\mbox{\scriptsize tot}}$
as
\begin{align}
w_{6}&=(0,0,0,1),\quad
w_{7}=(1,0,0,0), \nonumber \\
w_{10}&=(-1,2,-1,-1),\quad
w_{14}=(0,-1,3,0),            \\ 
w_{j}&=y_{0}:=(0,0,1,0), \quad j=8,9,11,12,13 \nonumber    \\
w_{i}&=y_{1}:=(0,1,0,0),\quad 1\leq i \leq 5  \nonumber 
\end{align}
The vector $y_{0}$ corresponds to 
the crepant divisor,
while the vector $y_{1}$, 
to a non-crepant divisor.

We find that for  a generic choice of $(r_{1},..,r_{4})$,
$M_{r}$ is a smooth 4-fold without Calabi-Yau property.

To see this, we can take the following charge assignment of $M_{r}$:
%suitable for the case
%
%
\begin{equation}
\begin{matrix}
 p_{1} & p_{2} & p_{3} & p_{4} & p_{5} & p_{8} & p_{9} &
 p_{11} & p_{12} & p_{13} & p_{6} & p_{7} & p_{10} & p_{14} & 
\mbox{\footnotesize FI}\\
%%%%%%%%%%%%
 1 &  0 &  0 &  0 &  0 &  -3&  0 &  
 0 &  0 &  0 &  0 &  0 &  0 &  1 & 
-r_{1}-r_{3}-2r_{4}\\
%%%%%%%%%%%%%%%%%%
 -2&  0 &  0 &  0 &  0 &  1 &  0 &  
 0 &  0 &  0 &  1 &  1 &  1 &  0 & 
r_{2}+r_{3}+r_{4}\\
%%%%%%%%%
 -1&  1 &  0 &  0 &  0 &  0 &  0 &  
 0 &  0 &  0 &  0 &  0 &  0 &  0 & 
r_{2}+r_{3}+r_{4}\\
 -1&  0 &  1 &  0 &  0 &  0 &  0 &  
 0 &  0 &  0 &  0 &  0 &  0 &  0 & 
-r_{1}\\
 -1&  0 &  0 &  1 &  0 &  0 &  0 &  
 0 &  0 &  0 &  0 &  0 &  0 &  0 & 
r_{2}\\
 -1&  0 &  0 &  0 &  1 &  0 &  0 &  
 0 &  0 &  0 &  0 &  0 &  0 &  0 & 
r_{2}+r_{3}\\
%%%%%%%%%%%%%%%%%%%%%%%%%%%%%
 0 &  0 &  0 &  0 &  0 &  -1&  1 &  
 0 &  0 &  0 &  0 &  0 &  0 &  0 & 
r_{2}\\
 0 &  0 &  0 &  0 &  0 & -1 &  0 &  
 1 &  0 &  0 &  0 &  0 &  0 &  0 & 
-r_{1}-r_{3}-r_{4}\\
 0 &  0 &  0 &  0 &  0 & -1 &  0 &  
 0 &  1 &  0 &  0 &  0 &  0 &  0 & 
-r_{1}-r_{4}\\
 0 &  0 &  0 &  0 &  0 & -1 &  0 &  
 0 &  0 &  1 &  0 &  0 &  0 &  0 & 
-r_{4}.\\
\end{matrix}\label{toricdata1112/5}
\end{equation}

It is easy to see from this toric data,
that $M_{r}$ is a smooth non-Calabi-Yau manifold
with Euler number $7$
for $r_{1}<0$, $r_{2}>0$, $r_{4}<0$,
$r_{2}+r_{3}+r_{4}>0$ and $r_{1}+r_{3}+r_{4}<0$.

The above observation remains valid even when $(r_{i})$ is restricted to
the subspace spanned  by the VEV of the bulk modulus field $\phi_{1}$,
$$
(r_{1},r_{2},r_{3},r_{4})=
t_{1}\left( 1,0,-1,-\frac{1}{2}(\sqrt{5}-1) \right)+
t_{2}\left( 0,1,\frac{1}{2}(\sqrt{5}-1),-\frac{1}{2}(\sqrt{5}-1) \right).
$$

%
%
%\newpage
\section{Discussion}
We have shown  in some examples that
the Higgs moduli space $M_{r}$ 
is a smooth 4-fold which is a blow-up of the original spacetime $X_{\Gamma}$
for a generic choice of the Fayet-Iliopoulos parameters
on the world volume.\footnote{Note that for B models 
we have not fully exploited enhanced (0,4) supersymmetry;
we may also have complex parameters associated with the F-term constraint
\cite{DM}, which we have set to zero in this article.}

At the same time, however, we have also shown
that  $M_{r}$ is a Calabi-Yau 4-fold only if 
$X_{\Gamma}$ admits a crepant blow-up to it,
in striking contrast with ${\Bbb C}^{2}/\Gamma$ or
 ${\Bbb C}^{3}/\Gamma$ cases.

Here we mention  some remaining  problems.
 
First though we have been mainly interested in the Higgs moduli space
in this article, it is important to
investigate the whole low energy effective theory of 
the world volume (0,2) gauge theory,
which is usually considered to flow to a chiral CFT in the IR limit.
It is also necessary to elucidate the holomorphic vector bundle 
$E$ on $M_{r}$ to which the massless part of 
the world volume left-handed fermions $\Lambda^{\mu}$ couple.

Second it would be interesting to consider the cases
in which $\Gamma\in \mbox{SU}(4)$ is not cyclic nor Abelian,\cite{SI3,JM}  
and the cases in which 
 the action of $\Gamma$ on the Chan-Paton factor 
$\gamma :\Gamma \mapsto \mbox{GL}(N,{\Bbb C})$ is not the 
regular representation.\cite{DGM}

\vfill

\begin{flushleft}
{\it Acknowledgement}
\end{flushleft}
We would like to thank N. Ishibashi for helpful discussions.

%%%%%%%%%%%%%%%%%%%%%%%%%%%%%%%%%%%%%%%%%%%%%%%%%%%%%%%%%%%%%%%%%%%%%%
\newpage
\appendix
\section{(0,2) Superformalism}
\noindent
Here we give the (0,2) convention used in this article.\cite{Phases}
See also \cite{Distler} for a different formulation.

A capital symbol (e.g. $Z$)
represents   a $n \times n$ matrix-valued field
$Z:=(z_{ij})$.\\
% its $(i,j)$ component denoted  by   $z_{ij}$.\\
We use $\lambda$ and $\psi$ for left-handed and right-handed fermions
respectively. \\ 
Gauge covariant superderivatives are defined by
\begin{equation}
\D= \frac{\partial}{\partial\th}-2i \thb \cov_{+},\quad
\Db = -\frac{\partial}{\partial\thb}+2i \th \cov_{+},
\end{equation}
where $\cov_{\pm}$ is the  gauge covariant derivative 
$
\cov_{\pm}Z=\partial_{\pm}Z+i[A_{\pm},Z]. \\
$
The remaining superpartners of the gauge multiplet appear in the operator 
\begin{equation}
\cov_{-}^{\mbox{\scriptsize s}}=\cov_{-}+(\th\overline
{\Lambda}_{\mbox{\scriptsize v}}
 +\thb\l_{\mbox{\scriptsize v}}+i\th\thb D).
\end{equation}
They form the basic field strength  multiplet:
\begin{equation}
\boldsymbol\Upsilon
:=\left[\cov_{-}^{\mbox{\scriptsize s}},\Db \right]
=\l_{\mbox{\scriptsize v}}+\th(2F_{+-}-iD)+2i\thb\th
\cov_{+}\l_{\mbox{\scriptsize v}}.
\end{equation}
A bosonic chiral superfield $\boldsymbol\Phi$,
which satisfies
$
\Db \boldsymbol\Phi=0,
$
is expanded  as follows,
\begin{equation}
\boldsymbol\Phi=Z+\sqrt{2}\th \Psi+2i\thb\th \cov_{+}Z,
\end{equation}
while the expansion of a fermionic chiral superfield 
$\boldsymbol\Lambda$ is 
\begin{equation}
\boldsymbol\Lambda=\l-\sqrt{2}\th G-2i\thb\th \cov_{+}\l
-\sqrt{2}\thb \boldsymbol\Xi,
\end{equation}
where $\boldsymbol\Xi$ is a bosonic chiral superfield and
$
\Db \boldsymbol\Lambda=\sqrt{2}\boldsymbol\Xi, \quad \Db \boldsymbol\Xi=0.
$

\section{Example of Anomaly Polynomial}
The symmetric matrix and the anomaly polynomial 
of  $\displaystyle{\frac{1}{4}}(1,1,1,1)$ model are 
\begin{equation}
C_{S}=
\begin{pmatrix}
\phantom{-}1  & -2  & \phantom{-}3  & -2  \\
-2  &  \phantom{-}1  &-2  &  \phantom{-}3  \\
 \phantom{-}3  & -2  & \phantom{-}1  & -2  \\
-2  &  \phantom{-}3  &-2  &  \phantom{-}1
\end{pmatrix},
\end{equation}
\begin{align}
{\cal A} &= (F_{1}^2+F_{2}^2+F_{3}^2+F_{4}^2)
-4(F_{1}F_{2}+F_{2}F_{3}+F_{3}F_{4}+F_{4}F_{1})
+6(F_{1}F_{3}+F_{2}F_{4})  \nonumber \\
&= +8\tilde{F}_{2}\tilde{F}_{2}
-2\tilde{F}_{1}\tilde{F}_{1}^{*}-2\tilde{F}_{3}\tilde{F}_{3}^{*},
\end{align}
where 
\begin{align*}
\tilde{F}_{1}&=\frac{1}{2}(iF_{1}-F_{2}-iF_{3}+F_{4}),\\
\tilde{F}_{2}&=\frac{1}{2}(-F_{1}+F_{2}-F_{3}+F_{4}),\\
\tilde{F}_{3}&=\frac{1}{2}(-iF_{1}-F_{2}+iF_{3}+F_{4}).
\end{align*}

%\newpage
%\section{Primitive Generators}

%------------------------------------------------------------------------

%
\end{document}